\documentclass[12pt]{mn2e}
\usepackage{graphicx}
\usepackage{rotating}

\newcommand{\civ}{C{\sc iv}}
\newcommand{\ciii}{C{\sc iii}]}
\newcommand{\mgii}{Mg{\sc ii}}

\title[New lensed quasars]
{New lensed quasars from the MUSCLES survey}
\author[Jackson et al.]
{Neal Jackson$^{1}$, Hayden Rampadarath$^{1,*}$, Eran O. Ofek$^{2,3}$, Masamune Oguri$^{4}$, 
Min-Su Shin$^{5}$\\
$^{1}$Jodrell Bank Centre for Astrophysics, School of Physics \& Astronomy, 
University of Manchester, Turing Building, Oxford Road, Manchester M13 9PL\\
$^{2}$Division of Physics, Mathematics and Astronomy, California Institute of
Technology, Pasadena, CA 91125, USA\\
$^{3}$Einstein Fellow\\
$^{4}$Institute for the Physics and Mathematics of the Universe, University 
of Tokyo, 5-1-5 Kashiwanoha, Kashiwa, Chiba 277-8583, Japan\\
$^{5}$Department of Astronomy, University of Michigan, 500 Church Street, Ann Arbor, 
MI48109-1042, USA\\
}
\input psfig.tex
\begin{document}
\maketitle
\begin{abstract}
Gravitational lens systems containing lensed quasars are important as 
cosmological probes, as diagnostics of structural properties of the lensing
galaxies and as tools to study the quasars themselves. The largest lensed
quasar sample is the SDSS Quasar Lens Search, drawn from the Sloan Digital
Sky Survey (SDSS). We are attempting to extend this survey using observations of 
lens candidates selected from a combination of the quasar sample from the SDSS 
and the UKIRT Infrared Deep Sky Survey (UKIDSS). This adds somewhat higher image quality
together with a wider range of wavelength for the selection process. 
In previous pilot surveys we observed 5
objects, finding 2 lenses; here we present further observations of 20 objects 
in which we find 4 lenses, of which 2 are independently discovered in SQLS 
(in preparation). Following earlier work on the combination of these two 
surveys, we have refined our method and find that use of a colour-separation 
diagnostic, where we select for separations between components which appear 
to decrease in wavelength, is an efficient method to find lensed quasars 
and may be useful in ongoing and future large-scale strong lensing surveys 
with instruments 
such as Pan-STARRS and LSST. The new lenses have mostly high flux ratios, with 
faint secondaries buried in the lensing galaxy and typically 6-10 times
less bright than the primary. Our survey brings the total number of lenses 
discovered in the SDSS quasar sample to 46, plus 13 lenses already known. 
This is likely to be up to 60-70\% of the total number of lensed quasars; 
we briefly discuss strategies by which the rest might be found.
\end{abstract}

\begin{keywords}
gravitational lensing - cosmology:galaxy formation
\end{keywords}

\footnote{Current address: International Centre for Radio Astronomy 
Research, Curtin University, GPO Box U1987, Perth, WA 6845, Australia}
\vskip -40mm
\large

\section{Introduction}

The first strong gravitational lens system was discovered over 30 years ago (Walsh,
Carswell \& Weymann 1979) and consisted of a quasar, lensed into two images about
six arcseconds apart. Since then, over a hundred lensed quasars have been discovered
by a variety of techniques.

Lensed quasars have a number of important uses. The most widely-studied application 
has been their use in determining the Hubble constant $H_0$ using measurements of 
time delays together with mass models for the lensing galaxy (Refsdal 1964); to date 
about 20 time delays have been measured (see e.g. Kochanek \& Schechter 2004; 
Jackson 2007 for reviews). If $H_0$ is accurately known from other observations, 
the combination of a time delay and known cosmological world model allows
the overall mass profile of the lens to be determined, independently of the mass
sheet degeneracy (Gorenstein et al. 1988) which otherwise plagues such 
measurements. Lensed quasar light curves can also be used to study microlensing
by stars in the lensed galaxy, (for a review see e.g. Wambsganss 1994); 
and variability as a function of wavelength in lensed quasars is a powerful
tool for investigation of the structure of the quasars themselves. This approach
has recently been used to constrain the sizes and structures of quasar accretion 
disks (Poindexter, Morgan \& Kochanek 2008). Four-image quasar lens systems are
valuable as they can be used to study both dark and luminous substructures in 
the lensing galaxy (Mao \& Schneider 1998, Dalal \& Kochanek 2002, Kochanek \& 
Dalal 2004, Vegetti et al. 2011, Jackson et al. 2010, Nierenberg et al. 2011).
Finally, if statistically complete samples of 
quasar lenses can be compiled, their redshift and separation distributions  
are powerful probes of both cosmology and galaxy evolution (Ofek, Rix \& Maoz 2003, 
Chae \& Mao 2003, Oguri et al. 2008, Capelo \& Natarajan 2007,
Matsumoto \& Futamase 2008).

Quasar lenses have hitherto been discovered in surveys of varying degrees of size and 
completeness. The largest of these surveys is the SDSS Quasar Lens Search (SQLS; 
Oguri et al. 2006, Oguri et al. 2008, Inada et al. 2010) in which a selection of 
quasars from the SDSS quasar list (Schneider et al. 2010) have been surveyed and 40 
lens systems discovered to date\footnote{The current list is maintained at the SQLS
website http://www-utap.phys.s.u-tokyo.ac.jp/$\sim$sdss/sqls/lens.html} together with
a further 2 so far in the survey described here. 
There are in total 105783 spectroscopically confirmed SDSS quasars in Data Release 7 
(DR7), so a selection is made which targets apparently extended quasar images for 
observation as potential lenses. This is done either by morphological selection, for 
objects of separation less than 2\farcs5 which are not deblended by the SDSS pipeline
(York et al. 2000), 
or by colour selection for other objects. A variant of this method has been suggested 
by Kochanek et al. (2006) who proposed that all variable and extended objects in future 
surveys be targeted as potential lenses.

Unfortunately, it is difficult to make such surveys complete, as the initial finding
survey throws up a number of false positives which must be eliminated. A significant
source of contamination is chance alignment with the most common type of star, M 
stars. Because of the depth of both the SDSS and UKIRT Infrared Deep Sky Survey 
(UKIDSS), which contain quasars
fainter than 19th magnitude, and the large size of the parent population of quasars, 
a significant number of M stars are expected within a typical Einstein radius of a 
number of SDSS quasars. Another source of contamination is foreground galaxies which 
are not close enough or massive enough to multiply image the background quasar, but which
still require followup as they may contain a weak lensed image. These are significantly 
more difficult to eliminate from the search,
as M stars have well-defined spectral features whereas many galaxies do not, and in
the case of high-flux ratio lenses, the separation of a faint lensed image from the
lensing galaxy can be difficult. This is especially the case because typical lensing
geometries in two-image lens systems place the faint lensed image very close to the
galaxy.

Several methods can be used to decrease the level of false positives. Firstly, we can
use colour information to try and guess the nature of the secondary, and indeed SQLS
uses colour cuts, particularly in the blue, to discriminate against systems where
no quasar secondary is present -- but at the cost of also discriminating
against high flux-ratio lens systems. Secondly, high resolution is an important aid to
lens-finding, because the median Einstein radius of a lens system is less than the 
typical ground-based seeing disk, corresponding to a typical lens galaxy which 
consists of an early-type system of 2-3$L_*$ and $M\sim10^{12}M_{\odot}$. In both of 
these respects it is helpful to use the UKIDSS survey (Lawrence et al. 2007) 
as an initial selection tool, because it expands the 
wavelength coverage into the infra-red and because its typical image quality 
(0\farcs7) is better than the typical image quality of SDSS (1\farcs4). The combined
SDSS wavelength coverage, of 350-2200nm, covers a range in a typical lens system
of a $z=2$ quasar and $z=0.5$ galaxy from the blue end, below the 4000\AA\ break of
the galaxy, where the quasar is expected to dominate the flux density and the red
end, where the old stellar population of the lensing galaxy should dominate the flux.
The MUSCLES (Major UKIDSS-SDSS Cosmic Lens Survey) has in the past used such 
a combination to discover two lenses: ULAS~J234311.9-005034 (Jackson et al. 2008)
and ULAS~J082016.1+081216 (Jackson et al. 2009). 

Here we extend the search with
further observations. In section 2 we present our candidate selection
and observations, and in section 3 we discuss the results. Finally we discuss, in
section 4, the implications for future large surveys for quasar lenses.

\section{Candidate selection and observations}

\subsection{Selection from SDSS and UKIDSS}

Of the 105783 quasars from the SDSS DR7 (Schneider et al. 2010), 35356 are in 
the most recent footprint area of the UKIDSS, Data Release 8, and have images in 
at least one of the UKIDSS colours: Z, Y, J, H and K. UKIDSS uses the
UKIRT Wide Field Camera (WFCAM; Casali et al, 2007); the photometric
system is described in Hewett et al (2006), and the calibration is
described in  Hodgkin et al. (2009). The pipeline processing and science
archive are described in Irwin et al (2011, in prep) and Hambly et al
(2008). We ignore the Z band, as it has in general images of lower signal-to-noise than
the other bands.

The UKIDSS images of all 35356 quasars were inspected by eye in order to find cases
where the quasar image is extended. In principle, this can be done automatically
using extension and ellipticity measurements for single objects in the UKIDSS database, 
together with selection of larger-separation lens candidates consisting of two UKIDSS 
objects close to each other. It is important to examine such cases individually, 
however, as automatic fitting is often unable to distinguish well between quasars 
residing in host galaxies, particularly at low redshift, and lens systems; automatically
fitted ellipticity and extension parameters can then be used to check that nothing
has been missed.

During the process which led to the discovery of ULAS~J082016.1+081216, we 
identified an additional diagnostic, which arises due to the nature of quasar lens 
systems. If the quasar is lensed into a bright image A and a normally weaker 
image B, the lens galaxy G lies along the line
from A to B, but closer to B. In practice, G and B are unresolved by most ground-based
optical images. However, because G is redder than B, the apparent separation between 
A and the G/B complex appears to decrease with increasing wavelength (Jackson et 
al. 2009; Rampadarath 2010). This effect is clearly seen in both the MUSCLES lenses,
as well as a considerable number of SQLS lenses, and we use it here to help 
increase the efficiency of the survey. 

Since the extensions in both SDSS and UKIDSS images are barely resolved, we 
measure the separation in a three-step process. First, we specify the positions 
of the two components approximately by eye, and fit for the fluxes only using 
two Gaussian profiles with a width approximately
given by the PSF of the images. These fluxes are then used as initial inputs to 
a fit in which we allow the positions and fluxes of the centres of the
two components to vary. Finally, we use the outputs of this iteration as inputs
to another fitting iteration where all four parameters of each Gaussian (x- 
and y- position, flux, and width) are allowed to vary, subject only to the 
constraint that the two widths of the Gaussians should be equal\footnote{As these
objects are potential lens systems, one object may be more extended than the other. 
However, our main purpose is to locate the distance between the centroids of two
condensations, and the fixing of the width of each avoids giving too much
freedom to the fit.}. This procedure
was found to optimize the stability and uniformity of the fits, and to be robust
in application to different images. Nevertheless, it does not always work well,
usually for obvious reasons such as one of the images being extremely faint. The
error on the separation is determined by the differences between the output
positions in the second and third fitting iterations. This error is usually
bigger than the photon statistics alone would indicate, but is more likely to
be representative of the uncertainty in the fitting.

The search by eye alone yields a list of 947 candidates from the original sample of
35356. 213 objects were removed because of the available RA range during the 
observations, leaving 734; of these 348 were judged by eye to be of higher priority
for observation. The Gaussian-fitting procedure was applied to this latter list.

\subsection{WHT spectroscopic observations}

Observations were made on the nights of 24 and 25 February 2011 using the ISIS
double-arm spectrograph on the 4.2-m William Herschel Telescope on La Palma. All 
observations used a 300 lines~mm$^{-1}$ grating on the red arm, and a 158 lines~mm$^{-1}$ 
grating on the red arm. The slit width was changed to attempt to match the seeing, which
varied from 0\farcs7 to 2\farcs0 during the observations, and the spectral resolution
obtained therefore varies considerably from object to object. With the median
slit of about 1\arcsec\ the spectral resolution was about 8\,\AA\ in the red arm
and 4\,\AA\ in the blue. The plate scale was 0\farcs22\,pix$^{-1}$ in the red arm 
and 0\farcs20\,pix$^{-1}$ in the blue arm. For each object,
a nearby offset star was used for acquisition, and in most cases a spectrum of the
offset star was taken in order to measure the spatial profile of the system at 
a location close to the target. Relative flux calibration was achieved using 
the standard star SP0501+527 (Oke et al. 1995), 
and the wavelength scale was set using observations of
Ne/Ar lamps once per night. All red-arm observations were centred on 730~nm, and
blue-arm observations on 430~nm, allowing complete coverage of the visible spectrum
from the atmospheric cutoff at 320~nm up to 900~nm, where the system sensitivity
and night-sky lines cut off the useful spectral range. All objects were observed
with a slit oriented along the position angle of the structure observed in
the UKIDSS images.

Objects were selected for observation by eye, from the candidate list. Preference
was given to objects which showed the separation-wavelength dependence outlined
above, and in fact all such objects were observed. Objects at z$>$0.5 were also
favoured, and objects at small separation were preferentially observed, as we know
from radio lens searches such as CLASS (Myers et al. 2003, Browne et al. 2003), 
which are sensitive to all separations down to
300~mas, that these are more likely to be lens systems. Table 1 presents a complete 
summary of the observations; a total of 20 objects were observed. Fig. 1 shows
the SDSS and UKIDSS images of the candidates.

\begin{figure*}
\begin{tabular}{cc}
{\bf 030214.8+000125} & \includegraphics[width=11.5cm]{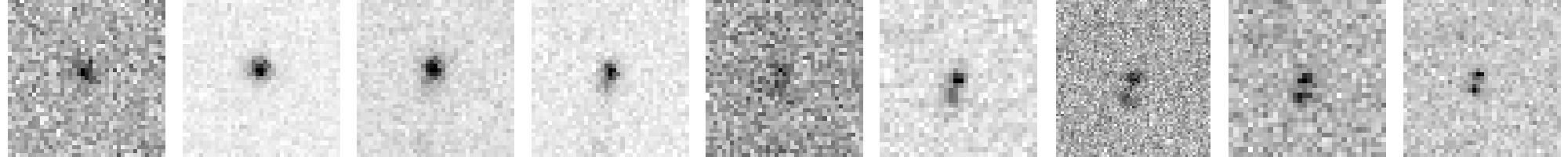}\\
{\bf 032732.1+041000} & \includegraphics[width=11.5cm]{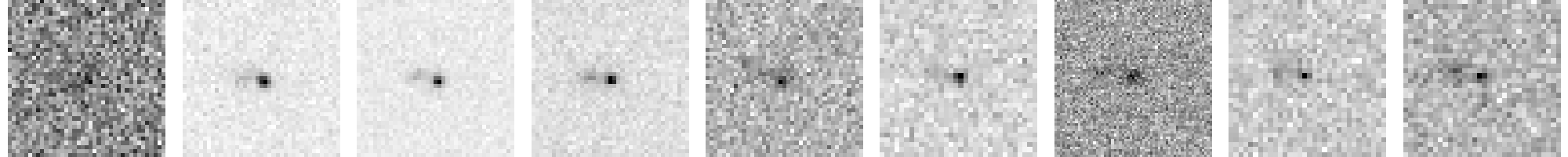}\\
{\bf 033248.5-001012} & \includegraphics[width=11.5cm]{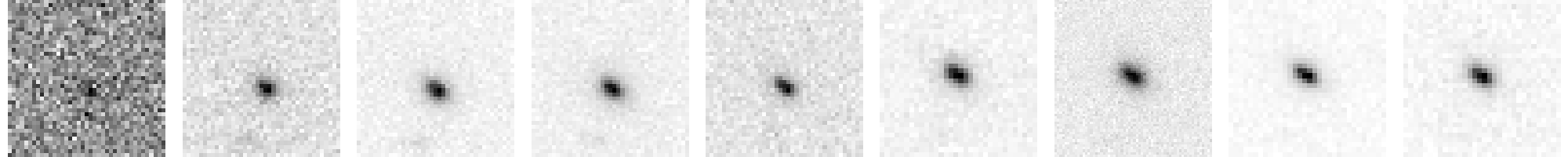}\\
{\bf 074352.6+245744} & \includegraphics[width=11.5cm]{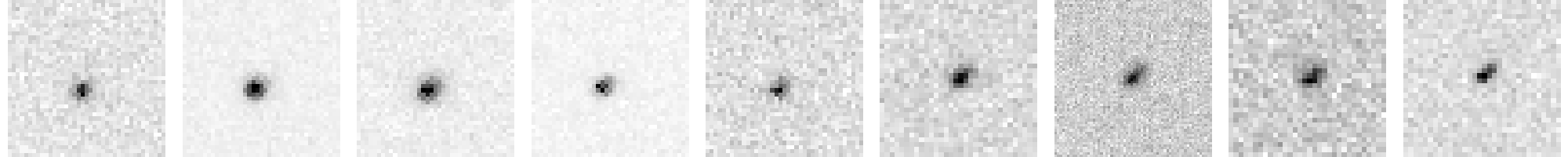}\\
{\bf 074557.0+282023} & \includegraphics[width=11.5cm]{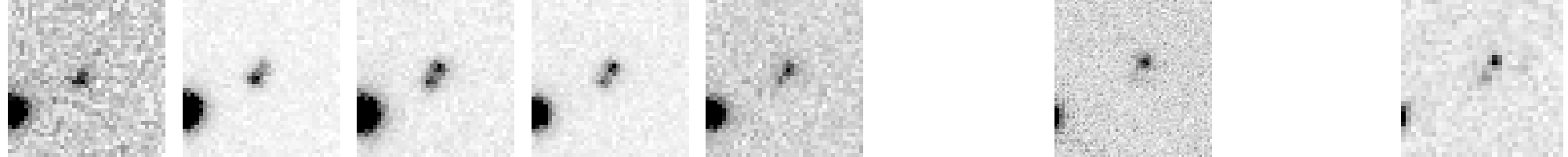}\\
{\bf 075901.3+284703} & \includegraphics[width=11.5cm]{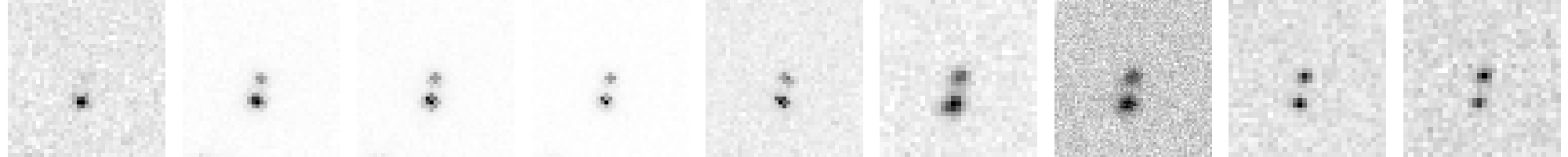}\\
{\bf 081910.2+211740} & \includegraphics[width=11.5cm]{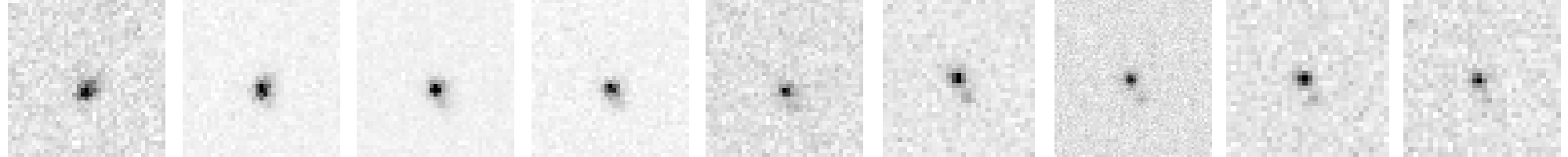}\\
{\bf 091750.5+290137} & \includegraphics[width=11.5cm]{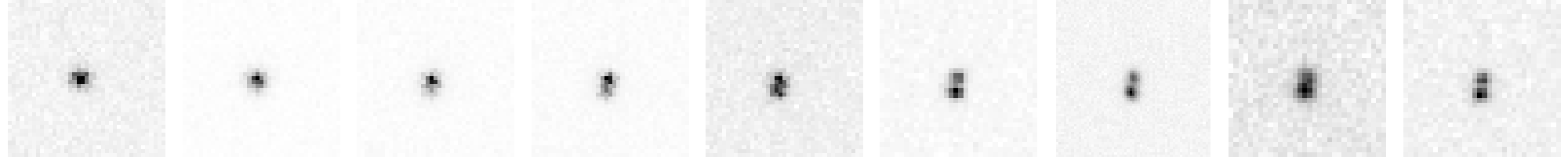}\\
{\bf 091831.6+110653} & \includegraphics[width=11.5cm]{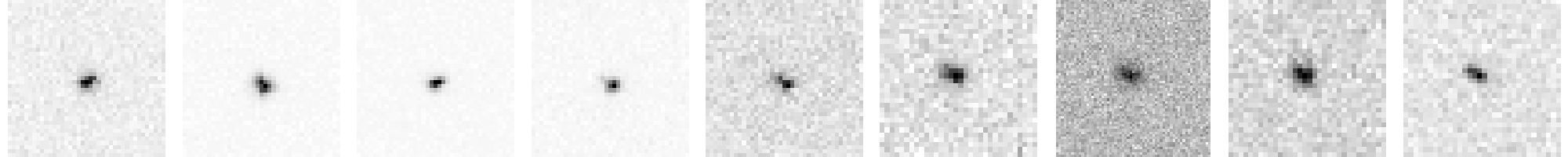}\\
{\bf 093850.0+055521} & \includegraphics[width=11.5cm]{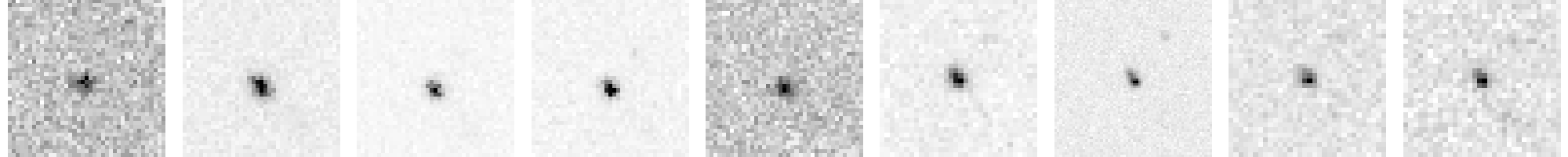}\\
{\bf 101105.7+061917} & \includegraphics[width=11.5cm]{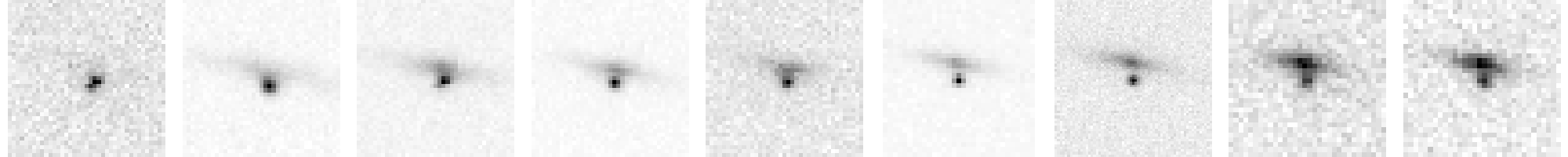}\\
{\bf 104141.7+091821} & \includegraphics[width=11.5cm]{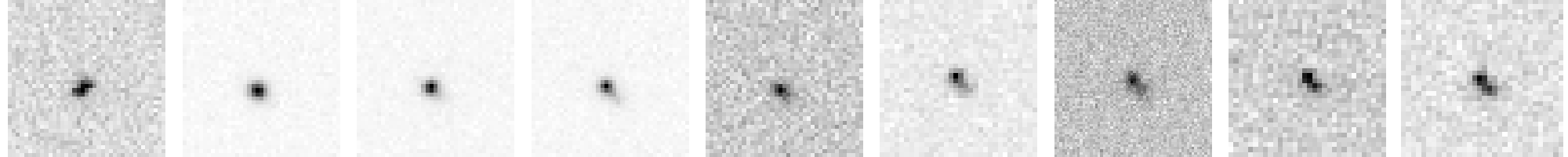}\\
{\bf 112342.8+104645} & \includegraphics[width=11.5cm]{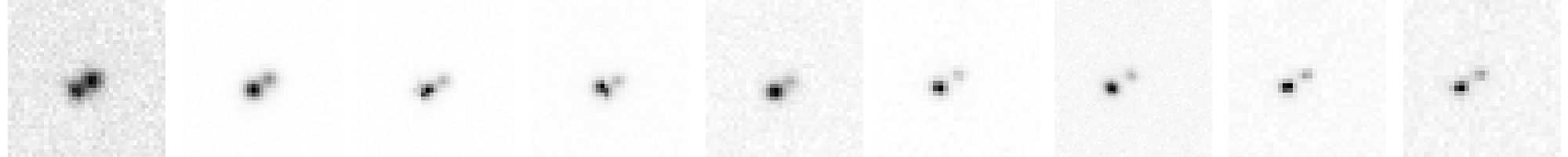}\\
{\bf 140515.4+095931} & \includegraphics[width=11.5cm]{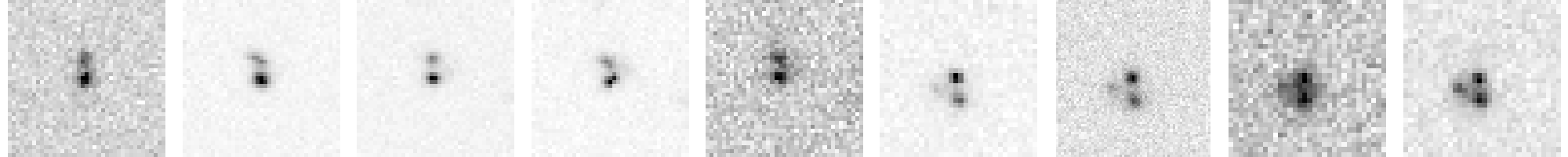}\\
{\bf 150824.2-000604} & \includegraphics[width=11.5cm]{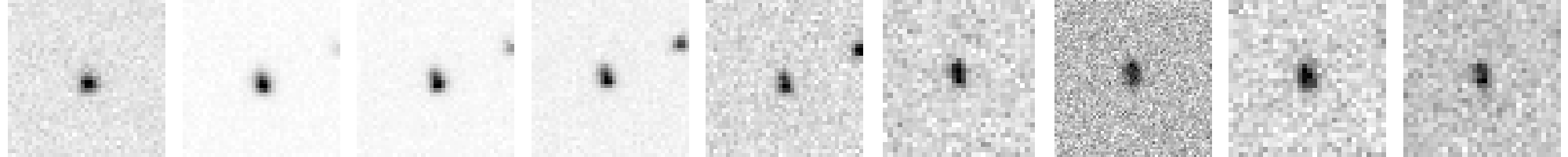}\\
{\bf 151543.0+025719} & \includegraphics[width=11.5cm]{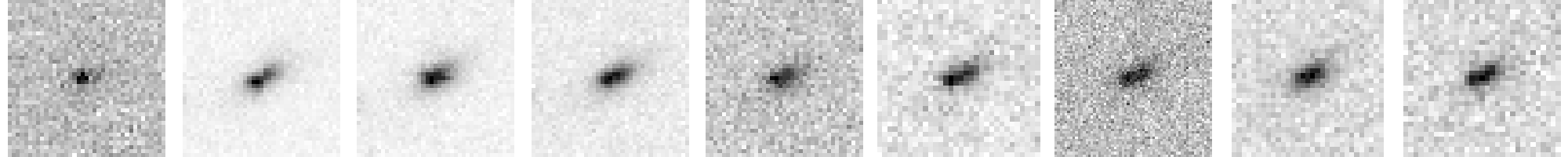}\\
{\bf 152625.4+014858} & \includegraphics[width=11.5cm]{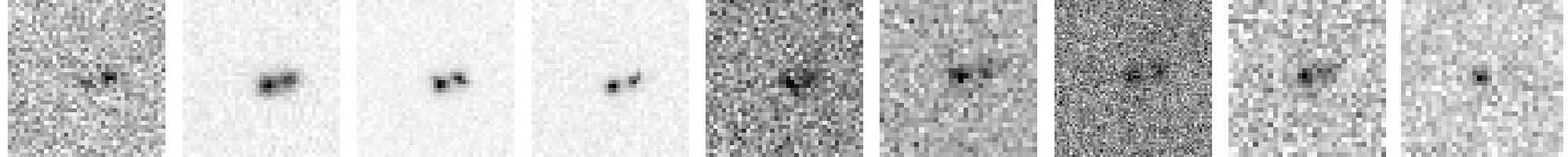}\\
{\bf 152720.1+014140} & \includegraphics[width=11.5cm]{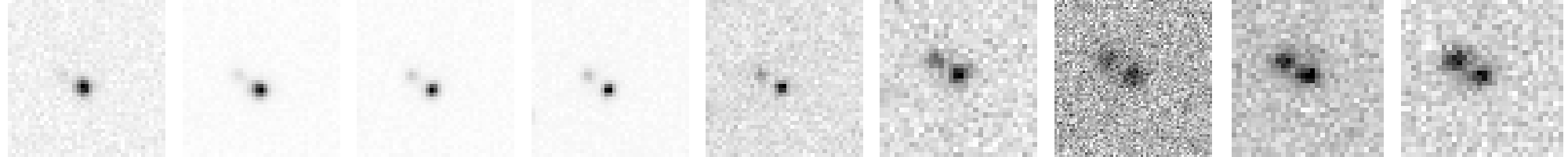}\\
{\bf 152938.9+103804} & \includegraphics[width=11.5cm]{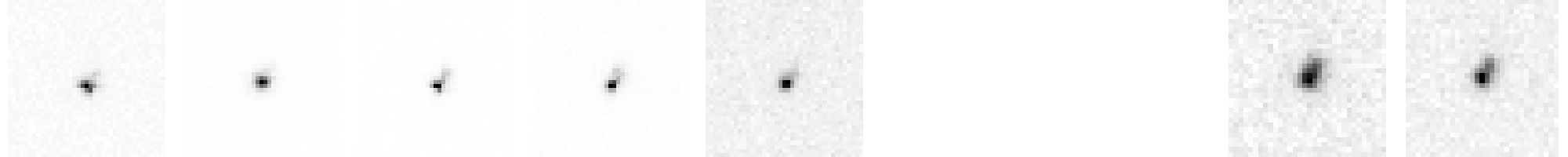}\\
{\bf 155105.3+261819} & \includegraphics[width=11.5cm]{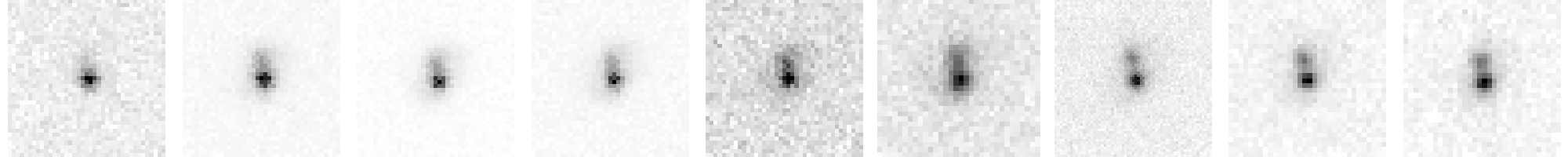}\\
\end{tabular}
\caption{Objects observed, as imaged by SDSS and UKIDSS. Columns are $u$, $g$, $r$,
$i$, $z$, $Y$, $J$, $H$, $K$ bands respectively.}
\end{figure*}

The WHT data were reduced using a mixture of the NOAO {\sc iraf} package and other 
routines written in Python. Second-order polynomial solutions for wavelength against 
pixel number were derived from the arc exposures and applied to the data, straightening
the spectral images in the wavelength direction in the process. The offset star
exposures were used to derive a distortion correction which was then used to 
straighten each part of the object spectra into the same spatial pixel. Finally,
the background was fitted and subtracted along the spatial direction in each column.


\subsection{APO imaging observations}

Two objects, 152720.1+014140 and 152938.9+103084, 
were observed with the SPIcam on the Astrophysical Research Consortium
3.5-meter telescope at the Apache Point Observatory on 2011 May 26.
We binned the 2048$^2$ CCD by a factor of two along each axis during the readout,
leading to a pixel scale of $0\farcs28$. The field of view is $4\farcm8$.
For each object, we obtained three dithered exposures of 180~s in both
$g$- and $i$-band under the seeing of 0\farcs9. The images were reduced using 
standard {\sc iraf} tasks.

\section{Results}

Inspection of the spectral images allowed immediate rejection of four candidates
as potential lens systems, because the secondary was obviously an M star;
these stars contain characteristic absorption bands at around 700~nm. This leaves
16 objects in which the secondary is likely to be a galaxy, either the host galaxy
of the quasar, a lensing galaxy, or an association along the line of sight which
is not close enough to lens the background quasar.

The diagnostic of a lens system is that the quasar emission lines are extended.
In order to test this, we first fitted the spatial profile of a region of the
spectral image which contains only the quasar continuum, using two Gaussians and 
using the width of the offset star's spatial profile to effectively measure the
seeing and thus guide the fit to the 
quasar. This provided an accurate position on the chip, in the spatial direction, 
for the quasar and the nearby galaxy. Next, continuum regions around a quasar 
emission line were interpolated across the emission line and subtracted off each 
spatial row on the chip. This left only photons from the emission line. The 
emission-line region was then extracted in the spatial direction and fitted again, 
using the spatial positions of the quasar and galaxy which were previously 
determined in the continuum fit. A possible detection is considered to have 
been made if the flux within the secondary Gaussian is greater than five times
the error on the fit.

\begin{table*}
\begin{tabular}{cccccccc}
Object & $z_{\rm SDSS}$ & RA (J2000) & Dec (J2000) & Sep (\arcsec) & $T_{\rm exp}$(min) & slit PA & ID\\
&&&&&\\
030214.8+000125 & 1.179 & 03 02 14.823 & +00 01 25.35 & 1.5 & 30 & 355 & G\\ 
032732.1+041000 & 3.110 & 03 27 32.113 & +04 10 00.09 & 2.1 & 45 & 80 & G\\ 
033248.4$-$001012 & 0.310 & 03 32 48.497&$-$00 10 12.37 & 1.1 & 25 & 60 & H\\ 
074352.6+245743 & 2.167 & 07 43 52.619 & +24 57 43.64 & 1.2 & 40 & 135 & L\\ 
074557.0+282023 & 1.558 & 07 45 57.015 & +28 20 23.23 & 1.5(z) &45 & 141 & G\\ 
075901.2+284703 & 2.849 & 07 59 01.287 & +28 47 03.42 & 2.2 & 10 & 350 & S\\ 
081910.1+211739 & 1.467 & 08 19 10.189 & +21 17 39.54 & 1.6 & 45 & 200 & G\\ 
091750.5+290137 & 1.816 & 09 17 50.529 & +29 01 37.47 & 1.3 & 40 & 175 & S\\ 
091831.5+110653 & 1.643 & 09 18 31.589 & +11 06 53.05 & 1.2(z) & 60 & 240 & L?\\ 
093849.9+055520 & 1.983 & 09 38 49.991 & +05 55 20.83 & 1.0 & 45 & 202 & L?\\ 
101105.6+061917 & 1.946 & 10 11 05.672 & +06 19 17.25 & 1.4 & 45 & 176 & G\\ 
104141.7+091820 & 1.068 & 10 41 41.729 & +09 18 20.55 & 1.6 & 75 & 40 & G\\ 
112342.7+104645 & 1.635 & 11 23 42.769 & +10 46 45.25 & 1.8 & 20 & 124 & S\\ 
140515.4+095931 & 1.810 & 14 05 15.421 & +09 59 31.30 & 1.8 & 45 & 0 & L\\ 
150824.2$-$000603 & 1.578 & 15 08 24.221&$-$00 06 03.85 & 1.1 & 45 & 10 & G\\ 
151543.0+025719 & 2.034 & 15 15 43.015 & +02 57 19.35 & 1.5 & 55 & 300 & H\\ 
152625.3+014857 & 2.987 & 15 26 25.389 & +01 48 57.95 & 2.1 & 45 & 280 & S\\ 
152720.1+014139 & 1.439 & 15 27 20.130 & +01 41 39.60 & 2.4 & 75 & 60 & L\\ 
152938.9+103803 & 1.971 & 15 29 38.902 & +10 38 03.90 & 1.5 & 45 & 330 & L\\ 
155104.1+261819 & 0.241 & 15 51 04.160 & +26 18 26.00 & 1.8 & 15 & 5 & H\\ 
\end{tabular}
\caption{Objects observed in the programme. Coordinates and redshifts are
from the SDSS. The final column (ID) represents the verdict of the analysis:
S=quasar+star, H=quasar+host galaxy, G=quasar+probably unassociated galaxy,
L=gravitational lens, L?=possible gravitational lens. Separations are measured
using the UKIDSS $Y$-band images, except for two cases where this fit fails
and the SDSS $z$-band is used instead.}
\end{table*}

We consider each gravitational lens, or possible gravitational lens, in turn.
These are the objects in which the fitting procedure described produces
evidence for the presence of extended emission lines. There are four objects
with fairly certain identification as lens systems, and two where the identification
is doubtful and requires further confirmation.

\subsection{ULAS~074352.6+245743} 

ULAS~074352.6+245743 contains a $z$=2.167 quasar, and the UKIDSS images reveal
a small separation (about 1\farcs0). The apparent separation between the fitted
components is noticeably smaller in the infrared than in the optical, although some 
of the SDSS images are difficult to fit due to the object being barely resolved.
The spectral fitting procedure has been applied to \ciii, because \civ\ is of lower
signal-to-noise and further into the blue where the seeing is worse, and
\mgii\ is badly contaminated by skylines. Fitting gives a 10$\sigma$ detection 
of an extension
to the line (Fig. 2) at a ratio of 6.4:1 between primary and secondary. The
galaxy is detected at low signal-to-noise, and only at wavelengths of
$>$550~nm. This suggests a redshift of a few tenths, although no absorption
line is visible which might help to confirm this. This object has independently
been identified as a likely lens by SQLS (unpublished; Oguri, private communication).

\begin{figure*}
\includegraphics[width=18cm]{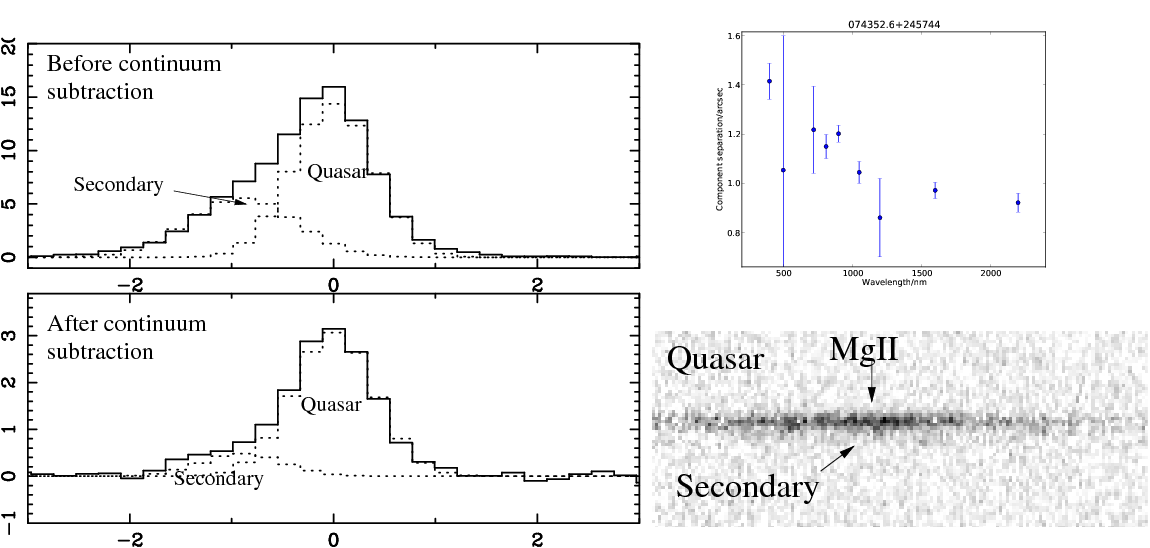}
\caption{ULAS~074352.6+245743. Bottom right: part of the continuum-subtracted 
spectral image around Mg{\sc ii}, showing a slight extension to the line. Top right: 
separation of the two components, as fitted
to the SDSS (optical) and UKIDSS (infra-red) images, showing the apparently smaller
separation in the infra-red due to the increased prominence of the lensing galaxy.
Top left: fit to the spatial profile of the continuum around C{\sc iii}], showing a 
clear extension due 
to either a companion galaxy or a lensing galaxy. Bottom left: spatial profile fit
to the line part of the spectral image, now with the continuum subtracted row by row.
Note that the raw profile is clearly asymmetric. A fit to the profile, with
the same centroids as determined earlier, suggests an extended line of
16\% of the flux of the primary line. }
\end{figure*}

\subsection{ULAS~140515.4+095931} 

\begin{figure*}
\includegraphics[width=18cm]{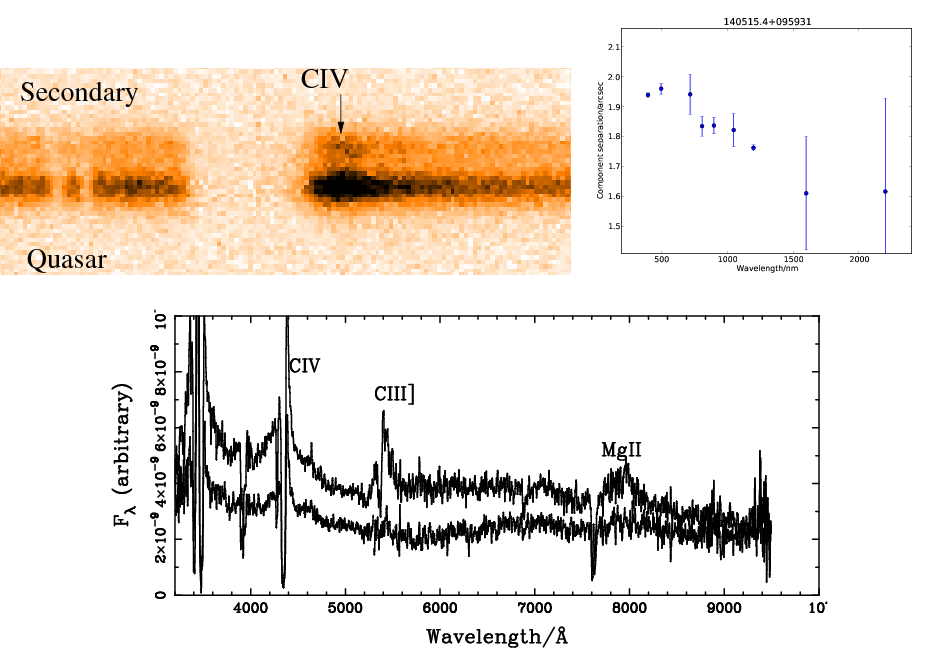}
\caption{ULAS~140515.4+095931. Top left: part of the spectral image around
Mg{\sc ii}, showing the double line. Top right: separation of the two components, as fitted
to the SDSS (optical) and UKIDSS (infra-red) images, showing the apparently smaller
separation in the infra-red due to the increased prominence of the lensing galaxy.
Bottom: WHT spectra, where the $x$-axis is observed-frame wavelength,
of the primary and secondary objects. The lens is a 1.7:1 double. The A and B 
atmospheric bands have been blanked.}
\end{figure*}

This system, containing a quasar of redshift 1.810, was selected for observation 
because of a noticeable
apparent decrease in separation with wavelength of its two components. It appears as
a double system in most optical wavebands, but has an additional component to
the east in the infrared. During the observations, the spectrograph slit was
positioned N-S along the two brightest components. These are separated by
1\farcs95 (Fig. 3) with emission lines clearly present in both
components. Spectra reveal an approximately 2:1 ratio in flux between the
$z=1.81$ quasar components, with very prominent associated absorption in both \civ\
and Ly$\alpha$. The continuum colours of the two condensations are different,
and consistent with the fainter quasar component being associated with a lensing
galaxy. The lensing galaxy itself displays a likely 4000\AA\ break, and possible
emission lines at 655 and 660~nm; although these are of low signal-to-noise and
require confirmation, this would imply an identification of CaH,K and a redshift
of 0.66 for the lensing galaxy. Like ULAS~074352.6+245743, this object has independently
been identified as a likely lens by SQLS (unpublished; Oguri, private communication).

\subsection{ULAS~152720.1+014139} 

This is a clear example of a new lens system. It consists of two images of
a $z$=1.439 quasar, separated by 2\farcs4. The spectrum around all of the major 
emission lines shows evidence of emission lines being clearly present
in the secondary. The object moved along the slit between individual
exposures in this object, but in all the exposures the two objects are seen
with the same separation.

The spectrum (Fig. 4) shows the \civ, \ciii\ and \mgii\ lines in both objects, 
with a flux ratio of 7.0$\pm$0.5:1. This is a relatively high ratio, similar to 
that obtained in many CLASS gravitational lenses. The two condensations in
the optical and infrared images have very different colours, due to the
weakness of the secondary lensed quasar image and consequent dominance of the
lens galaxy over it at all but the bluest wavelengths. The different colours
also immediately rule out the hypothesis that we are dealing with a binary
quasar system as opposed to a gravitational lens. No absorption lines are
present which might suggest the redshift of the galaxy, but the pronounced
upturn in the spectrum between 500 and 600~nm is likely to be the 400~nm
break feature of an early-type galaxy, leading to a probable redshift of
about 0.3 for the lens galaxy. This system shows an apparent decrease in 
separation of the two components with wavelength; all of the UKIDSS images 
show a separation between the two components about 10\% smaller than the 
SDSS optical images.

Fig. 4 also shows the APO $i$-band image of the system. This has been fitted 
using the {\sc galfit} package (Peng et al. 2002) with two models; either two point
sources, with PSFs determined from a nearby unsaturated star in the field, or two 
point sources plus a S\'{e}rsic profile to represent a possible lensing galaxy. The
residuals from each fit show clearly that an extended object is required close
to the secondary component, which we identify as the lensing galaxy (Table 2). The fit
implies a flux ratio between the point sources of 9.7$\pm$0.9, slightly greater 
than that inferred from spectroscopy, and a separation of 2\farcs7$\pm$0\farcs1. 
The galaxy is fitted by a S\'{e}rsic profile of index 3.1 and effective radius of 
0\farcs43, slightly elongated ($b/a$=0.66) in PA 13$^{\circ}$. Because of the 
limited signal-to-noise, the fitted parameters of this galaxy are likely to be 
degenerate with each other, in particular the effective radius and S\'{e}rsic index.

\begin{figure*}
\includegraphics[width=18cm]{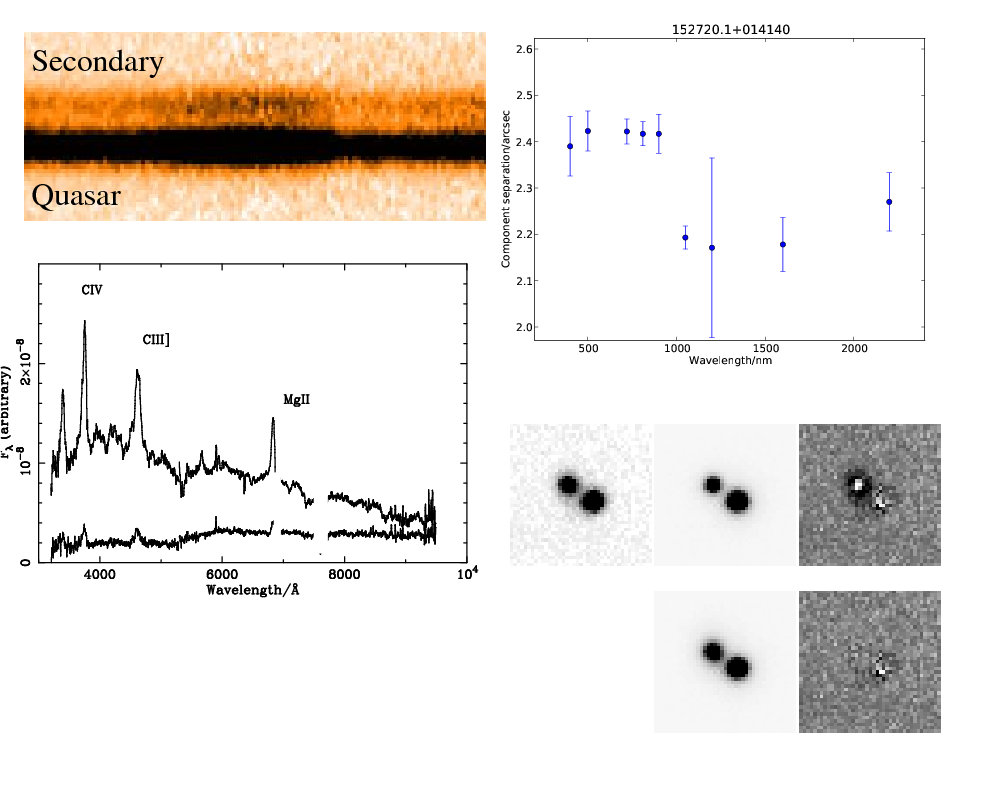}
\caption{ULAS~152720.1+014139. Top left: part of the spectral image around
Mg{\sc ii}, showing the double line. Top right: separation of the two components, as fitted
to the SDSS (optical) and UKIDSS (infra-red) images, showing the apparently smaller
separation in the infra-red due to the increased prominence of the lensing galaxy.
Bottom left: WHT spectra, where the $x$-axis is observed-frame wavelength,
of the primary and secondary objects. The kinks in the
spectrum at 5300\,\AA are the effects of the dichroic cut. The A and B atmospheric
bands have been blanked. Bottom right: APO image
(left) together with model and {\sc galfit} residual with two point sources (top) and
two point sources plus a S\'{e}rsic-profile galaxy (bottom).}
\end{figure*}

\begin{table*}
\begin{tabular}{lccccc}
Object & Relative & Relative & S\'{e}rsic & Axis ratio & $r_e$ \\
&centroid&flux&index&($b/a$)& (arcsec)\\
152720.1-Q1 & (0,0) & 1.00 & - & - & - \\
152720.1-Q2 & ($-2.20$,1.38) & 0.10 & - & - & - \\
152720.1-G  & ($-1.86$,1.24) & 0.35 & 3.1 & 0.66 & 0.43 \\
152938.9-Q1 & (0,0) & 1.00 & - & - & - \\
152938.9-Q2 & (0.69,1.18) & 0.17 & - & - & - \\
152938.9-G  & (0.52,0.77) & 0.17 & 2.0$^*$ & 0.5$^*$ & 0.30 \\
\end{tabular}
\caption{Parameters of GALFIT fits to the images of ULAS~152720.1+014139 and
ULAS~152938.9+103803. Centroids in arcseconds (N,W) of the bright quasar
image are given in column 2, and fluxes relative to the bright quasar
image in column 3. A star indicates that this parameter has been fixed at 
a reasonable value in order to stabilise the otherwise badly constrained
fit (see text). In the case of ULAS~152938.9+103803, the uncertainties on
fluxes and positions of the weaker components are high (probably about 0.2
arcseconds in position). In both cases, S\'{e}rsic index and effective radius
are degenerate.}
\end{table*}

\subsection{ULAS~152938.9+103803} 


This $z=1.97$ quasar clearly has a secondary close to it. Fitting
around the \mgii\ line, as previously described, yields a secondary
in the emission line with a flux ratio of 4.9$\pm$0.5 between
the primary and secondary (Fig. 5) and a notional significance of
$10\sigma$. Fitting to the \ciii\ line, after subtraction, also
shows an extended line, although here the continuum is more difficult
to subtract. The blue frame also shows a likely extension to the
\civ\ line after continuum subtraction (Fig. 5). 
On the SDSS and UKIDSS images, there is a possible decrease 
in separation with wavelength, although the errors are high and
some fits fail altogether due to the faintness of the secondary. The
separation of the components on the spectra is 1\farcs2, although this
is probably an underestimate because of the presence of the
lensing galaxy; on the SDSS images we see a separation of about 1\farcs5. 

These initial tests on the spectroscopic data are not conclusive, because
the effects are relatively small in this object and may be subject to
errors in the continuum fitting, as well as possible guiding errors.
We therefore performed further tests, this time on the Ly$\alpha$
line in the blue spectrum. Each of the three independent spectral images
from the three exposures with the ISIS blue arm was straightened
so that the spectrum runs horizontally on the chip, and the continuum
was again subtracted using a straight-line fit in each spatial pixel
to areas blueward and redward of the line. For each frame, we then
examined the spatial profile of the prominent absorption line in the centre
of Ly$\alpha$, together with the adjacent line emission (Fig. 5). Both the
absorption line and emission line spatial profiles show asymmetry in
the same direction, in all three frames. The fact that all three frames
behave similarly suggests that telescope guiding errors are not the cause
of the asymmetry in the spatial profiles. The fact that the asymmetry
appears in the same direction in emission and absorption suggests that 
mis-subtraction of the continuum
is not responsible, as an under-subtraction in the emission part of the line
would correspond to an over-subtraction in the absorption part. This
result should be robust for any type of mis-subtraction, including no
subtraction of the continuum at all. We therefore conclude that the spectra 
provide evidence for an asymmetric profile of line emission, and hence the 
presence of a secondary image of the quasar.

The APO i-band images of this system (Fig.5) show the two main components
clearly. Again, fitting using {\sc galfit} to the image is difficult using
only two PSFs. Inclusion of a S\'{e}rsic profile causes the $\chi^2$ to fall
by about 30\% over the whole image, and a residual peak close to the fainter
image to be removed. The fitting (Table 2) also implies a separation of about 
1\farcs3 
between the two point components. However, the three-component fit is not well 
constrained due to the relatively low signal-to-noise; this free fit 
results in an extremely elliptical galaxy ($b/a=0.06$) but good removal of 
the remaining residual. Constraining the galaxy to be more circular improves
the residual by slightly less, but suggests a flux ratio of about 4--5 between the
image components, close to that derived from spectroscopy. In both cases the
effect of including a galaxy is to remove an extended region in the residual
close to the weaker image.

\begin{figure*}
\includegraphics[width=18cm]{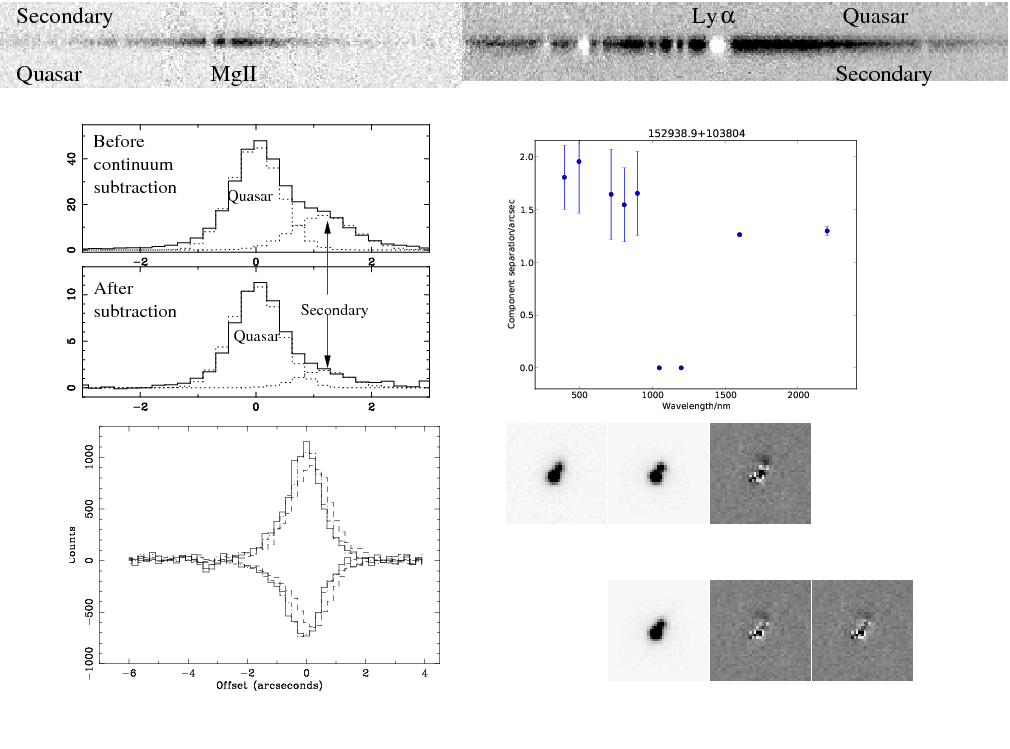}
\caption{ULAS~152938.9+103803. Top: part of the continuum-subtracted 
spectral image around \mgii (left) and Ly$\alpha$ (right), showing in each case 
a slight extension to the line. Middle left: separation of the two components
in the spatial profile of the Mg{\sc ii} line, at the top without subtraction
of the continuum and at the bottom with continuum subtraction; a small residual
remains, indicating the presence of extended line emission. The fit suggests
extended line emission of about 17\% of the primary.
Middle right: fit to the separation of two components in the
SDSS (optical) and UKIDSS (infra-red) images, showing the apparently smaller
separation in the infra-red due to the increased prominence of the lensing galaxy.
The fit fails in two cases just above 1000~nm. 
Bottom left: Spatial profile fits to the continuum subtracted Ly$\alpha$ 
emission line, from the three exposures separately (solid, dashed and 
dash-dotted lines), both for the prominent region of absorption and also
for the adjacent emission region. All six profiles yield asymmetries, suggesting
that the asymmetry observed is not due to mis-subtraction of the continuum
or to guiding errors. Bottom right: APO image (left) together with model and 
{\sc galfit} residual with two point sources (top) and two point sources plus 
a S\'{e}rsic-profile galaxy (bottom). In the second case, the residuals resulting 
from a free fit (left) and from a fit with $b/a$ fixed at 0.5 and the S\'{e}rsic 
index fixed at 2.0 are shown.}
\end{figure*}

\subsection{ULAS~091831.5+110653 and ULAS~093849.9+055520} 

Two further objects still remain as candidates, although their status is
uncertain. The first, ULAS~091831.5+110653, 
has a possible very slight extension on the SDSS images,
in a direction just north of west. On the UKIDSS images the extension is
more definite, and is likely to be much redder than the quasar itself. There
is marginal evidence for an apparent decrease of separation with wavelength,
although the $z$-image is probably the only SDSS image on which the
extension appears with enough signal-to-noise for the separation to be
determined with any accuracy. Fitting to the spectral image around the
\mgii\ line, with the continuum subtracted, gives a nominal 4$\sigma$
detection of an extension to the line, but this is not very convincing
given the uncertainties associated with the continuum subtraction and
the Gaussian fitting to spatial profiles. The seeing was unfortunately
not optimal during this observation, at about 1\farcs5, and this object, while 
still a candidate lens system, requires further confirmation.

%
%


ULAS~093849.9+055520 is a candidate small-separation ($\sim1\farcs0$)
lens system. Unfortunately, the spectrum was taken during 
a period of poor seeing, the object has weak and broad emission
lines, and the quasar redshift puts \mgii\ in a bad region of
OH skylines at around 830-840~nm. We attempted a fit of the 
line and continuum around \civ , but although it is possible
to derive an extended line profile, this is heavily dependent upon
the fitting of each spatial row to subtract the surrounding continuum.
This object, while still a candidate, therefore requires further
observations to confirm or rule out its lensing status.

\section{Discussion and conclusions}

Figure 6 shows an updated version of the relationship between separation
and magnitude difference of SQLS and MUSCLES gravitational
lens samples.
The figure also includes a comparison with the CLASS sample, which is smaller
but should be highly complete in both component flux ratio ($<$2.5 magnitudes)
and separation. Since the CLASS survey was selected using VLA observations 
at 8.4~GHz, which have a diffraction-limited resolution of 0\farcs22, 
all lens systems on scales of 0\farcs3 should be detected, with
some possibility of finding lenses on scales smaller than this.

About one-third of CLASS lenses have separations and flux ratios within the
SQLS statistical sample limits ($>$1\arcsec and $<$1.25 mag, Inada et al. 2010). 
Within this
range there are about 30 known lenses within the SDSS-DR7 quasars. Hence we expect
about 90 lens systems in total from the SDSS-DR7 quasars, corresponding to
a lensing rate of roughly 1:1000. The fact that 59 have been found to date
(40 in SQLS, 13 previously known, 6 in the MUSCLES survey) suggest that only
about 30 remain to be found. Those that remain are likely to be either of
large flux ratio, leading to difficulties in separating the weak component
from the lensing galaxy, or to be of small separation. Fig. 6 strongly 
suggests that a large majority of the lenses yet to be found in SDSS-DR7
are in the latter category. In principle, further research using the UKIDSS
lens candidate systems can help to find the remaining objects, although the
success rate is likely to drop. This is because the separation-wavelength
diagnostic that we have developed is unlikely to be useful below about
1 arcsecond separation, as such systems will be completely unresolved
on most SDSS images.

\begin{figure}
\includegraphics[width=6.5cm,angle=-90]{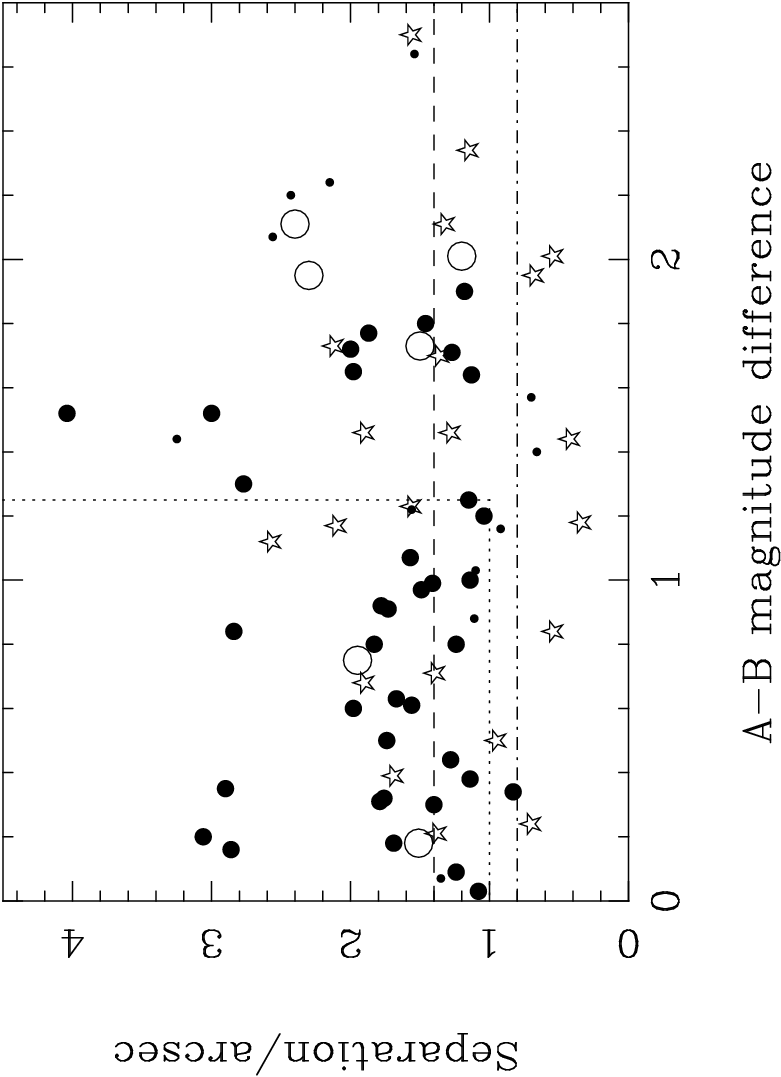}
\caption{Updated plot of separation vs. magnitude difference between
the brightest component and the next brightest component of gravitational
lens systems (third brightest in the case of quad lenses). Large filled
circles represent SQLS lenses, and small filled circles represent SDSS
quasars already known before SQLS to be lens systems. Stars represent
CLASS lens systems, which should be complete to a separation of 300~mas
and a magnitude difference of 2.5. Open circles represent MUSCLES
lenses from previous papers together with this work. The UKIDSS median
image quality (dot-dashed line) and SDSS (dashed line) are indicated,
together with the dynamic range and lens separation limit of the SDSS
statistical sample (dotted line). Three SDSS lenses lie outside the plot
due to large separation.}
\end{figure}

We have investigated the selection function of this survey in more detail 
using a simulation of the selection process. A large sample of simulated
objects was generated for a range of primary-secondary flux ratios up
to 2 magnitudes, and for a range of separations between 0 and 3 arcseconds.
A magnitude for the primary object was chosen from the distribution of
H-band magnitudes of SDSS quasars from the Schneider et al. (2010) catalogue,
which is approximately described by a mean of
17.4 and a standard deviation of 0.78. For each simulated object, a random
decision was taken by the algorithm on whether or not to include the 
secondary object, with equal probability of each outcome. A blind test 
was then undertaken for 2000 objects in which an observer (NJ) judged whether 
or not the quasar was extended, with the correct decision being scored
as 1 and an incorrect decision as 0. Because half the simulated objects
were lenses, by chance an average score of 0.5
would be expected, with 1.0 representing perfect discrimination in the
selection process. In Fig. 7 we show the results of these
simulations, binned according to flux ratio and separation. It is clear
that the selection function involves both variables, with a very high
probability of success for separations $>1\arcsec$ and flux ratios less
than one magnitude, but with significant probability of success at 
slightly smaller separations and higher flux ratios. The simulation
was then repeated with 20000 objects, using the {\sc sextractor} package 
(Bertin \& Arnouts 1996) for selection of extended objects. In this 
simulation, objects were deemed to be extended if more than one object 
was detected by {\sc sextractor}, or if a single object with major-to-minor 
axis ratio $>r$ was fitted, with values of $r$ of 1.3 and 1.5. The results 
are similar to those obtained by eye (Fig. 7). In principle, lower values
of $r$ give more efficient selection, although in practice, lowering
$r$ will increase vulnerability to false positives caused by effects
not simulated here such as variable PSFs.

In fact, UKIDSS
provides four images at different colours, which effectively increase
the efficiency of selection of faint secondary images. To include this
effect, the simulation was repeated with four images being generated
for each field using different random noise. Fig. 7 (right panel) shows the
results of this procedure, and it is now clear that $>$80\% efficiency
is achieved in selection for separation $>1\arcsec$ and flux ratios
$<$1.3 magnitudes, but with significant efficiency at or beyond 2.0
magnitudes. Unsurprisingly, the separation limit at just under 1$\arcsec$ is
less affected by the extra information.

\begin{figure*}
\begin{tabular}{cc}
\includegraphics[width=8cm]{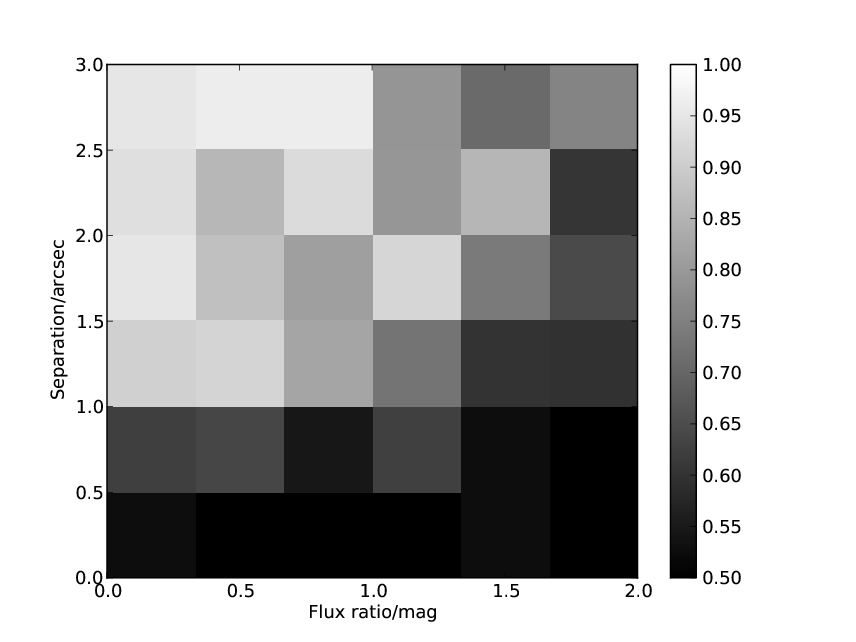}&
\includegraphics[width=8cm]{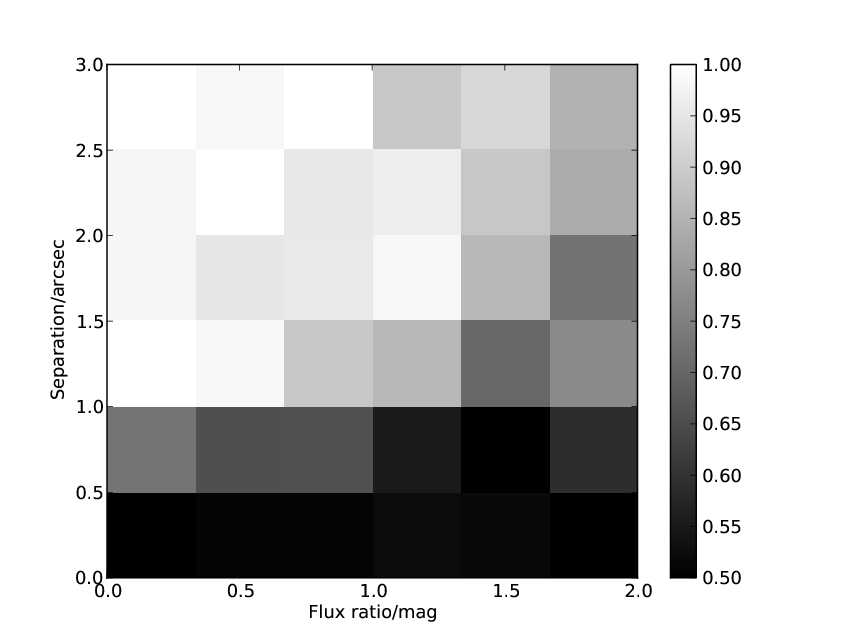}\\
\includegraphics[width=8cm]{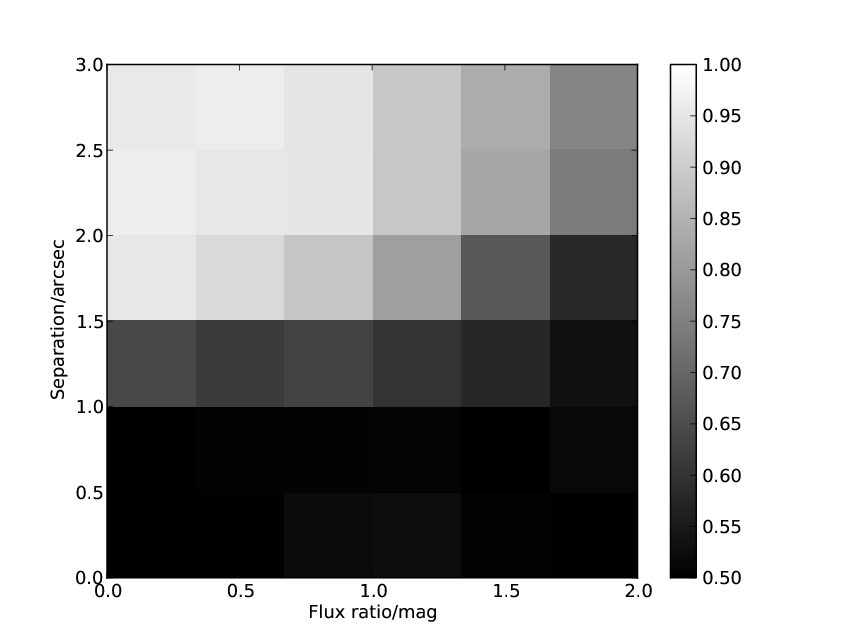}&
\includegraphics[width=8cm]{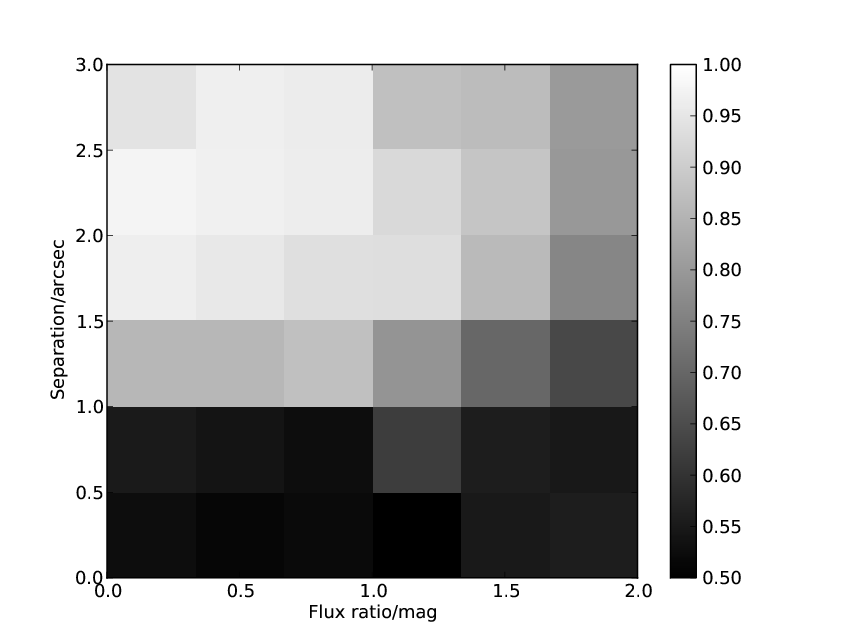}\\
\end{tabular}
\caption{Fraction of correct results in simulations of lensed quasars
as a function of separation and flux ratio, for a sample similar to 
UKIDSS (see text). A value of 0.5 is expected by chance, whereas perfect
selection gives a value of 1.0. Top left: results obtained using one image 
and selection by eye. 
Top right: results obtained using four similar images, corresponding
to the case for UKIDSS where images in four colours are available.
Bottom: results obtained using one image and
selection by {\sc sextractor} (see text) using $r$=1.5 (left) and $r$=1.3 (right). }
\end{figure*}

Looking at Figs. 6 and 7 together, it is probable that a few SDSS quasar
lenses remain at high flux ratios, and that these lie in the half of the
sky not yet fully covered by the MUSCLES programme. Such objects may
well have been excluded from SQLS because of the colour criterion, with
the secondary colour being dominated by the lensing galaxy. The majority
(20-25 lenses) are likely to lie at separations less than 0\farcs8 - 0\farcs9,
in the region where CLASS is much more efficient because of its three times
higher resolution.

The results presented here have implications for quasar surveys using future
generations of telescopes. In the near term, the Pan-STARRS survey (Kaiser 
et al. 2002), which is already operating, is expected to complete surveys
of large areas of the sky with resolutions of about 1\arcsec ; 
similar to, or a little better than, SDSS. Further into the future, the LSST 
project (Ivezic et al. 2008) expects to observe the entire southern hemisphere 
to much greater depth in multiple visits, with still better image quality
(about 0\farcs7). This will find a very large number of strong lens systems 
(Oguri \& Marshall 2010) but from a finding sample too large to rely on
complete searches by eye, as has hitherto been possible (e.g. Jackson 2008) 
and automated algorithms will be required. In either case, isolation of complete 
samples of quasar lens systems in an automatic fashion based on imaging data alone 
is not a straightforward process, and is strongly dependent on image quality.

\section*{Acknowledgements}
We thank Michael Strauss and Ian Browne for comments on the paper. 
The WHT is operated on the island of La Palma by the Isaac Newton Group of Telescopes 
at the Spanish Observatorio del Roque de los Muchachos of the Instituto de 
Astrof\'{\i}sica de Canarias. This work also uses data from the Apache Point 
Observatory 3.5-metre telescope, which is owned and operated by the Astrophysical 
Research Consortium. We would like to thank the Kavli Institute for 
Theoretical Physics and the organizers of the KITP workshop ``Applications of 
Gravitational Lensing'' for hospitality. This work began at this KITP workshop.
HR was supported by the EU under the Marie Curie Early-Stage Training
network MEST-CT-2005-19669 ``Estrela''. EOO is supported by an Einstein Fellowship
and NASA grants. 
This work was supported in part by the National Science Foundation under grant no. 
PHY05-51164 and by the Department of Energy contract DE-AC02-76SF00515. This
work is based on data obtained as part of the UKIRT Infrared Deep Sky Survey, 
UKIDSS (www.ukidss.org). This work was supported in part by the FIRST program 
"Subaru Measurements of Images and Redshifts (SuMIRe)", World Premier 
International Research Center Initiative (WPI Initiative),MEXT, Japan, and 
Grant-in-Aid for Scientific Research from the JSPS (23740161).
Funding for
the SDSS and SDSS-II has been provided by the Alfred P. Sloan
Foundation, the Participating Institutions, the National Science
Foundation, the US Department of Energy, the National Aeronautics and
Space Administration, the Japanese Monbukagakusho and the Max-Planck
Society, and the Higher Education Funding Council for England. The SDSS
web site is http://www.sdss.org/. The SDSS is managed by the
Astrophysical Research Consortium (ARC) for the Participating
Institutions.

\section*{References}

\small



\noindent Bertin, E., Arnouts, S., 1996, A\&AS, 117, 393




\noindent Browne I.W.A., et al., 2003,  MNRAS, 341, 13. 

\noindent Capelo P.R., Natarajan P. 2007, NewJPhys 9, 445 

\noindent Casali M., et al., 2007,  A\&A, 467, 777. 

\noindent Chae, K.-H., Mao, S., 2003, ApJ, 599, L61 

\noindent Dalal, N., Kochanek, C.S., 2002, ApJ 572, 25



\noindent Gorenstein,  M.V., Shapiro, I.I., Falco, E.E,, 1988, ApJ, 327, 693 

\noindent Hambly N.C., et al. 2008,  MNRAS, 384, 637. 

\noindent Hewett P.C., Warren S.J., Leggett S.K., Hodgkin S.T. 2006,  MNRAS, 367, 454. 

\noindent Hodgkin S.T., Irwin M.J., Hewett P.C., Warren S.J., 2009,MNRAS, 394, 675 


\noindent Inada N., et al., 2010, AJ, 140, 403.








\noindent Ivezic, Z., et al., 2008, Serbian Astron. J., 176, 1

\noindent Jackson N., 2008, MNRAS, 389, 1311

\noindent Jackson N., Ofek E.O., Oguri M. 2008,  MNRAS, 387, 741.  

\noindent Jackson N., Ofek E.O., Oguri M. 2009,  MNRAS, 398, 1423 

\noindent Jackson N., 2007, LRR, 10, 4 

\noindent Jackson N., Bryan S., Mao S., Li C., 2010, MNRAS, 403, 826


\noindent Kaiser N., et al., 2002, SPIE, 4826, 154  





\noindent Kochanek, C.S., Dalal, N., 2004, ApJ, 610, 69

\noindent Kochanek C.S., Schechter P., 2004, in ``Measuring and Modeling the Universe'',
Carnegie Obs. Cent. Symp., ed. Freedman W.L., Cambridge University Press, p. 117 

\noindent Kochanek, C.S., Mochejska,B., Morgan, N.D., Stanek, N.Z., 2006, ApJ, 637, L73 


\noindent Lawrence A., et al., 2007,  MNRAS, 379, 1599. 

\noindent Mao, S., Schneider, P., 1998, MNRAS, 295, 587



\noindent Matsumoto A., Futamase T. 2008,  MNRAS, 384, 843. 



\noindent Myers S.T., et al., 2003,  MNRAS, 341, 1. 

\noindent Nierenberg, A., Auger, M.W., Treu, T., Marshall, P.J., Fassnacht, C.D.,
2011, ApJ, 731, 44

\noindent Ofek E.O., Rix H.-W., Maoz D. 2003, MNRAS 343, 639 





\noindent Oguri M., et al., 2006, AJ, 132, 999   

\noindent Oguri, M., Marshall, P.J., 2010, MNRAS, 405, 2579



\noindent Oguri M., et al., 2008, AJ, 135, 512  

\noindent Oke J.B., et al., 1995, PASP 107, 375 

\noindent Peng C.Y., Ho L.C., Impey C.D., Rix H.-W., 2002, AJ 124, 266 



\noindent Poindexter S., Morgan N., Kochanek C.S., 2008, ApJ, 673, 34 

\noindent Rampadarath H., 2010, MSc thesis, University of Manchester 

\noindent Refsdal S. 1964,  MNRAS, 128, 307. 


\noindent Schneider D.P., et al., 2010, AJ, 130, 2360   

\noindent Vegetti S., Koopmans, L.V.E., Bolton, A.S., 
Treu, T., Gavazzi, R., 2011, MNRAS, 408, 1969

\noindent Wambsganss J., 1994, in Kochanek C.S., Schneider P., Wambsganss J., Proc. 33rd Saas-Fee
Advanced Course, ed. Meylan G., et al., Springer-Verlag, Berlin 

\noindent Walsh D., Carswell R., Weymann, R.J., 1979, Nature, 279, 381 

\noindent York D.G., et al., 2000, AJ, 120, 1579

\end{document}